\documentclass[preprint,showpacs,preprintnumbers,amsmath,amssymb]{revtex4}

\usepackage{graphicx}% Include figure files
\usepackage{dcolumn}% Align table columns on decimal point
\usepackage{bm}% bold math

\begin{document}

\preprint{}

\title{Two-Qubit 
Separabilities as Piecewise Continuous Functions of Maximal Concurrence. II--The Relevance of Dyson Indices}

\author{Paul B. Slater}% 
\email{slater@kitp.ucsb.edu}
\affiliation{%
ISBER, University of California, Santa Barbara, CA 93106\\
}%
\date{\today}% It is always \today, today,

\begin{abstract}
We importantly amend a certain parenthetical remark made
in Part I (arXiv:0806.3294), to the effect 
that although two-qubit {\it diagonal-entry}-parameterized separability
functions had been shown 
(arXiv:0704.3723) to clearly conform to a pattern dictated by 
the ``Dyson indices'' ($\beta = 1$ [real], 2 [complex],
4 [quaternionic])
of random matrix theory, 
this did not appear to be the case with 
regard to {\it eigenvalue}-parameterized separability functions (ESFs). 
But upon further examination of the extensive numerical analyses 
reported in Part I, 
we find quite convincing evidence that adherence to the Dyson-index 
pattern does also hold for ESFs,
at least as regards the 
upper {\it half}-range 
$\frac{1}{2} \leq C \leq 1$ of the maximal concurrence 
over spectral orbits, $C \equiv
C(\lambda_1\ldots\lambda_4)= \max \{0,\lambda_1 -\lambda_3 -2
\sqrt{\lambda_2 \lambda_4}\}$,
$\lambda_1 \geq \lambda_2 \geq \lambda_3 \geq \lambda_4$, with
the $\lambda$'s being the eigenvalues of associated 
$4 \times 4$ density matrices. 
To be specific, it strongly appears 
that in this upper half-range, the {\it real} two-qubit ESF is 
simply proportional
to $(2-2 C)^{\frac{3}{2}}$, and its {\it complex} counterpart--in 
conformity to the Dyson-index pattern--proportional
to the {\it square} of the real ESF, 
that is, $(2-2 C)^3$. The previously documented piecewise 
continuous (``semilinear'') behavior
in the {\it lower} half-range $0 \leq C \leq \frac{1}{2}$ still appears, however,  
to lack any particular Dyson-index-related interpretation.

\end{abstract}


\maketitle
Part I of this study  \cite{maxconcur4} had been devoted to the
question of determining for the generic 
(9-dimensional) real and (15-dimensional)
complex two-qubit systems, the nature of certain 
trivariate ``eigenvalue-parameterized separability
functions'' (ESFs). These 
(metric-{\it independent}) 
ESFs, it was argued, could substantially 
assist in the determination of
separability 
{\it probabilities} in terms of certain metrics (the Hilbert-Schmidt
and Bures being the most conspicuous examples).
We further investigated 
in \cite{maxconcur4} the possibility that these {\it prima facie} 
trivariate functions of the eigenvalues $\lambda_i$ $(i =1,\ldots 4)$ 
of $4 \times 4$ density matrices $(\lambda_4=1-\Sigma_i^3 
\lambda_i)$, were expressible as {\it univariate} 
functions 
\begin{equation} \label{ansatz}
S_4^{(\beta)}(\lambda_1\ldots\lambda_4) = \sigma^{(\beta)}
(C(\lambda_1\ldots\lambda_4)),
\end{equation}
of the {\it maximal concurrence} $C$ over
spectral orbits  \cite[sec. VII]{roland2} \cite{ishi,ver},
\begin{equation} \label{maxcon}
C(\lambda_1\ldots\lambda_4)= \max \{0,\lambda_1 -\lambda_3 -2
\sqrt{\lambda_2 \lambda_4}\},
\hspace{.2in} \lambda_1 \geq \lambda_2 \geq \lambda_3 \geq \lambda_4.
\end{equation}
(At this starting point in our presentation, let us regard 
$\beta$ in (\ref{ansatz}) 
only as a notational [dummy variable], not calculational device--motivated 
by Dyson-index conventions--taking
the values 1 [real], 2 [complex], 4 [quaternionic].)

Our main conclusions in \cite{maxconcur4} 
were that--if the reducibility-to-univariance property 
(\ref{ansatz}) held, as our extensive numerical evidence 
appeared to suggest could be the case--the 
associated real and complex 
univariate functions both had jumps of approximately
$50\%$ magnitude at $C=\frac{1}{2}$, as well as a number of additional
discontinuities (remarkably 
coincident in both the real and complex cases) in the 
{\it lower} half-range $C \in [0,\frac{1}{2}]$. Also, both univariate functions
appeared to be 
simply {\it linear} between certain of these discontinuities. 
The {\it upper} half-range $C 
\in [\frac{1}{2},1]$--in which the univariate functions 
of $C$ took lesser values--did 
not command our attention 
in \cite{maxconcur4}, seeming to be of relatively less 
interest. Our only pertinent observation there was that
there did not appear to be any discontinuities 
in this segment.

\begin{figure}
\includegraphics{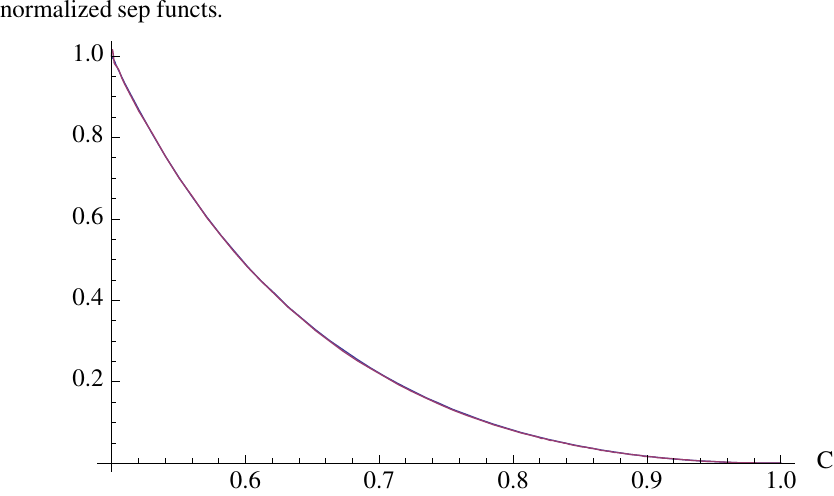}
\caption{{\it Joint} plot of numerical estimates of
$\Big( \frac{\sigma^{real}(C)}{\sigma^{real}(\frac{1}{2})} \Big)^2$ {\it and}
$ \frac{\sigma^{complex}(C)}{\sigma^{complex}(\frac{1}{2})}$ for 
$C \in [\frac{1}{2},1]$
\label{fig:normed}}
\end{figure}

\begin{figure}
\includegraphics{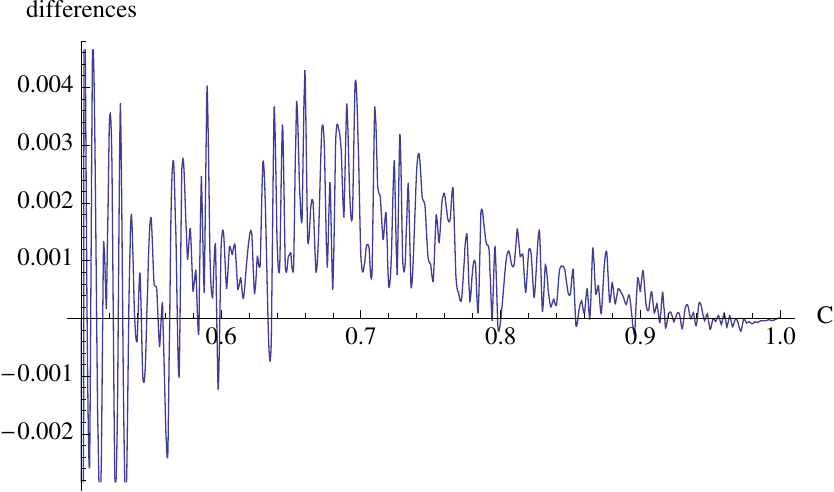}
\caption{\label{fig:newDyson}Numerical estimate of 
$\Big( \frac{\sigma^{real}(C)}{\sigma^{real}(\frac{1}{2})} \Big)^2 -
 \frac{\sigma^{complex}(C)}{\sigma^{complex}(\frac{1}{2})}$}
\end{figure}

\begin{figure}
\includegraphics{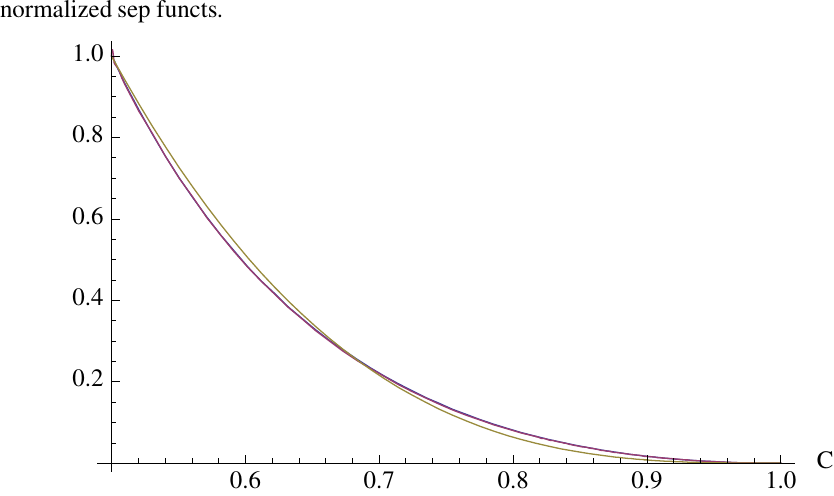}
\caption{\label{fig:normed2}The two functions in Fig.~\ref{fig:normed}, 
along with the 
additional ({\it very} 
closely-fitting) function $(2 -2 C)^3$}
\end{figure}

Now, in fact, 
turning our attention 
more closely to this upper half-range, we readily find
strong evidence for a very interesting Dyson-index-type phenomenon.
If we {\it normalize} our extensive 
numerical estimates from \cite{maxconcur4} 
of $\sigma^{1}(C)$ and  $\sigma^{2}(C)$
to both equal 1 at $C=\frac{1}{2}$, then a joint plot 
(Fig.~\ref{fig:normed}) of the latter 
normalized (complex) function versus the {\it square} of the former 
normalized 
(real) function for $C \in [\frac{1}{2},1]$ remarkably 
shows {\it no} perceptible difference between the two resulting curves. 
(The sample 
[quasi-Monte Carlo] estimate of 
$\sigma^{complex}(\frac{1}{2})$ is 0.0651586 and that of 
$\sigma^{real}(\frac{1}{2}))$ is 0.1803748.)
In Fig.~\ref{fig:newDyson},
we show--on a much finer scale than used 
in Fig.~\ref{fig:normed}--the actual (very small) differences 
\begin{equation}
\Big( \frac{\sigma^{real}(C)}{\sigma^{real}(\frac{1}{2})} \Big)^2 -
 \frac{\sigma^{complex}(C)}{\sigma^{complex}(\frac{1}{2})}
\end{equation}
between them.
Of further 
considerable importance, Fig.~\ref{fig:normed2} is a repetition of 
Fig.~\ref{fig:normed}, but along with the insertion now 
of the function
\begin{equation} \label{simple}
(2 -2 C)^3 = 8 (1-C)^3,
\end{equation}
which we see fits our two estimates {\it very} well.
Assuming that (\ref{simple}) is the correct form 
(up to the still 
not exactly-known normalization factor) of $\sigma^2(C)$ over 
$C \in [\frac{1}{2},1]$, we can estimate the 
associated contribution from density matrices corresponding to 
this half-range 
to the Hilbert-Schmidt and Bures separability probabilities of
generic complex two-qubit systems to be 0.041568 and 0.0267378, respectively.
(The real counterparts 
of these separability probabilities are, then, 
0.134611 and 0.104113, respectively.)

Let us further note that our sample estimate of
the ratio
\begin{equation} \label{ratio}
\frac{\sigma^{complex}(\frac{1}{2})}{\Big(\sigma^{real}(\frac{1}{2})\Big)^2}
= \frac{0.0651586}{0.1803748^2} = 2.00272
\end{equation}
is very close (and possibly theoretically exactly 
equal) to 2.

Over $0 \leq C \leq \frac{1}{2}$, the range of primary 
interest in \cite{maxconcur4}, the estimates of
the real and complex two-qubit separability functions {\it intersect} 
(near $C=0.1812$),
and appear to have linear segments 
over the {\it same} subintervals \cite[Figs. 1, 5, 7]{maxconcur4}. 
These features, of course, make any obvious application of
the Dyson-index pattern problematical in this half-range.
So, the behaviors of the univariate functions $\sigma^\beta(C)$, 
($\beta =1$ [real], 2 [complex]),
over the two indicated regimes of $C$ seem to be highly distinct.
The point $C=\frac{1}{2}$ clearly serves as a point of 
major behavioral transition, 
with the lower half-range now appearing 
perhaps to be the more theoretically
challenging of the two. An outstanding
question would seem to be what are the specific values of 
$\sigma^{real}(\frac{1}{2})$ and $\sigma^{complex}(\frac{1}{2})$, which 
we used as normalization factors in our analyses above. The
nearness to 2 of the ratio (\ref{ratio}) may be a helpful guide 
in this regard.

Our analyses of two-qubit {\it diagonal-entry}-parameterized separability
functions \cite{slaterPRA2,slater833,slaterJGP2} and 
{\it eigenvalue}-parameterized separability functions 
\cite{slaterJMP2008,maxconcur4} 
have shared a {\it common} goal: the determination of two-qubit separability
volumes and probabilities (in terms of various metrics). 
As pieces of these formidable objectives
begin to be assembled, we can pose a further challenge--to find
transformations between the two {\it different} 
sets of coordinates used--that is, 
(1) the diagonal entries and (2) the eigenvalues
of $4 \times 4$ density matrices--that will map one set of 
separability functions
into the other.
The Schur-Horn
Theorem, which asserts
that the decreasingly-ordered vector of eigenvalues  of an
Hermitian matrix {\it majorizes} the decreasingly-ordered vector of
its diagonal entries \cite[chap. 4]{hornjohnson} (cf. \cite{nielsenvidal}), 
would appear to be of possible relevance in this regard, particularly 
since the maximal concurrence $C$ over spectral orbits (\ref{maxcon}) 
is expressed
in terms of the {\it ordered} eigenvalues.

\begin{acknowledgments}
I would like to express appreciation to the Kavli Institute for Theoretical
Physics (KITP)
for computational support in this research.
\end{acknowledgments}

\bibliography{MaxRevise2}% Produces the bibliography via BibTeX.

\begin{thebibliography}{10}
\expandafter\ifx\csname natexlab\endcsname\relax\def\natexlab#1{#1}\fi
\expandafter\ifx\csname bibnamefont\endcsname\relax
  \def\bibnamefont#1{#1}\fi
\expandafter\ifx\csname bibfnamefont\endcsname\relax
  \def\bibfnamefont#1{#1}\fi
\expandafter\ifx\csname citenamefont\endcsname\relax
  \def\citenamefont#1{#1}\fi
\expandafter\ifx\csname url\endcsname\relax
  \def\url#1{\texttt{#1}}\fi
\expandafter\ifx\csname urlprefix\endcsname\relax\def\urlprefix{URL }\fi
\providecommand{\bibinfo}[2]{#2}
\providecommand{\eprint}[2][]{\url{#2}}

\bibitem[{\citenamefont{Slater}()}]{maxconcur4}
\bibinfo{author}{\bibfnamefont{P.~B.} \bibnamefont{Slater}},
  \eprint{arXiv:0806.3294 (to appear in J. Phys. A)}.

\bibitem[{\citenamefont{Hildebrand}(2007)}]{roland2}
\bibinfo{author}{\bibfnamefont{R.}~\bibnamefont{Hildebrand}},
  \bibinfo{journal}{J. Math. Phys.} \textbf{\bibinfo{volume}{48}},
  \bibinfo{pages}{102108} (\bibinfo{year}{2007}).

\bibitem[{\citenamefont{Ishizaka and Hiroshima}(2000)}]{ishi}
\bibinfo{author}{\bibfnamefont{S.}~\bibnamefont{Ishizaka}} \bibnamefont{and}
  \bibinfo{author}{\bibfnamefont{T.}~\bibnamefont{Hiroshima}},
  \bibinfo{journal}{Phys. Rev. A} \textbf{\bibinfo{volume}{62}},
  \bibinfo{pages}{022310} (\bibinfo{year}{2000}).

\bibitem[{\citenamefont{Verstraete et~al.}(2001)\citenamefont{Verstraete,
  Audenaert, and Moor}}]{ver}
\bibinfo{author}{\bibfnamefont{F.}~\bibnamefont{Verstraete}},
  \bibinfo{author}{\bibfnamefont{K.}~\bibnamefont{Audenaert}},
  \bibnamefont{and} \bibinfo{author}{\bibfnamefont{B.~D.} \bibnamefont{Moor}},
  \bibinfo{journal}{Phys. Rev. A} \textbf{\bibinfo{volume}{64}},
  \bibinfo{pages}{012316} (\bibinfo{year}{2001}).

\bibitem[{\citenamefont{Slater}(2007{\natexlab{a}})}]{slaterPRA2}
\bibinfo{author}{\bibfnamefont{P.~B.} \bibnamefont{Slater}},
  \bibinfo{journal}{Phys. Rev. A} \textbf{\bibinfo{volume}{75}},
  \bibinfo{pages}{032326} (\bibinfo{year}{2007}{\natexlab{a}}).

\bibitem[{\citenamefont{Slater}(2007{\natexlab{b}})}]{slater833}
\bibinfo{author}{\bibfnamefont{P.~B.} \bibnamefont{Slater}},
  \bibinfo{journal}{J. Phys. A} \textbf{\bibinfo{volume}{40}},
  \bibinfo{pages}{14279} (\bibinfo{year}{2007}{\natexlab{b}}).

\bibitem[{\citenamefont{Slater}(2008{\natexlab{a}})}]{slaterJGP2}
\bibinfo{author}{\bibfnamefont{P.~B.} \bibnamefont{Slater}},
  \bibinfo{journal}{J. Geom. Phys.} \textbf{\bibinfo{volume}{58}},
  \bibinfo{pages}{1101} (\bibinfo{year}{2008}{\natexlab{a}}).

\bibitem[{\citenamefont{Slater}(2008{\natexlab{b}})}]{slaterJMP2008}
\bibinfo{author}{\bibfnamefont{P.~B.} \bibnamefont{Slater}},
  \textbf{\bibinfo{volume}{doi:10.1016/j.geomphys.2008.08.008}}
  (\bibinfo{year}{2008}{\natexlab{b}}).

\bibitem[{\citenamefont{Horn and Johnson}(1991)}]{hornjohnson}
\bibinfo{author}{\bibfnamefont{R.~A.} \bibnamefont{Horn}} \bibnamefont{and}
  \bibinfo{author}{\bibfnamefont{C.~R.} \bibnamefont{Johnson}},
  \emph{\bibinfo{title}{Matrix Analysis}} (\bibinfo{publisher}{Cambridge
  Univ.}, \bibinfo{address}{New York}, \bibinfo{year}{1991}).

\bibitem[{\citenamefont{Nielsen and Vidal}(2001)}]{nielsenvidal}
\bibinfo{author}{\bibfnamefont{M.~A.} \bibnamefont{Nielsen}} \bibnamefont{and}
  \bibinfo{author}{\bibfnamefont{G.}~\bibnamefont{Vidal}},
  \bibinfo{journal}{Quant. Inform. Comput.} \textbf{\bibinfo{volume}{1}},
  \bibinfo{pages}{76} (\bibinfo{year}{2001}).

\end{thebibliography}

\end{document}